\documentclass[runningheads]{llncs}
\usepackage[T1]{fontenc}
\usepackage{amsmath}
\usepackage{amsfonts}
\usepackage{amssymb}
\usepackage{graphicx}
\usepackage{xcolor}
\usepackage{subcaption}
\usepackage[misc]{ifsym}
\usepackage{orcidlink}
	\renewcommand{\orcidID}[1]{\,\orcidlink{#1}}
\usepackage[nameinlink,capitalise]{cleveref}
\usepackage{complexity}

\newcommand{\bigO}{\mathcal{O}}

\DeclareMathOperator{\Gcal}{\mathcal{G}}
\DeclareMathOperator{\Hcal}{\mathcal{H}}
\DeclareMathOperator{\Lcal}{\mathcal{L}}

\DeclareMathOperator{\Rcal}{\mathcal{R}}

\DeclareMathOperator{\Wcal}{\mathcal{W}}
\DeclareMathOperator{\Xcal}{\mathcal{X}}

\DeclareMathOperator{\inter}{int} %
\DeclareMathOperator{\bd}{\partial} %

\DeclareMathOperator{\tw}{tw} %

\begin{document}
	\title{Complete Rectangulations are no Easier to Guard}
	\author{Vasco Cruz%
		$^{\textrm{(\Letter)}}$
		\orcidID{0000-0003-4263-9382}
		\and
		Ana Paula Tomás
		\orcidID{0000-0002-1776-4897}}
	\authorrunning{V.\,Cruz and A.\,P.\,Tomás}
	\institute{CMUP, Faculdade de Ciências, Universidade do Porto,\\Rua do Campo Alegre s/n, 4169--007 Porto, Portugal\\
		\email{vasco.cruz@fc.up.pt, aptomas@fc.up.pt}}
	\maketitle
	\begin{abstract}
		A rectangulation is a partition of a rectangular domain $\Omega$ into $n$ pairwise interior-disjoint rectangles such that no four of them meet at a common point.
		Nearly two decades ago, Dinitz, Katz and Krakovski [EWCG 2007; IJCGA 2009] showed that finding a minimum vertex-guard set for a partial rectangulation with holes is $\NP$-complete, leaving open whether the problem remains hard for complete rectangulations (without holes) and whether a $4/3$-approximation can be improved.
		In this paper, we answer both questions in the affirmative.
		We first prove that vertex-guarding rectangulations (\textsc{RVG}) is $\NP$-complete even when every rectangle must be guarded, and even when a fixed subset of vertices is forbidden.
		We then obtain a PTAS for \textsc{RVG} by formulating the problem as a red-blue dominating set on graphs of bounded outerplanarity,
		thus ruling out $\APX$-hardness unless $\P = \NP$.
		Finally, we describe a family of boundary-exposed rectangulations, in which every rectangle touches the outer boundary $\bd \Omega$.
		We establish structural properties of their weak duals and guillotine partitions, and show that \textsc{RVG} on these rectangulations is solvable in linear time.
		This approach extends to rectangulations with at most $k$ interior rectangles, yielding a fixed-parameter tractable algorithm with running time $2^{\bigO(k)} \cdot n$.
		
	\end{abstract}

	\section{Introduction}
	
\begin{figure}[tb]
		\centering
		\includegraphics[scale=0.3]{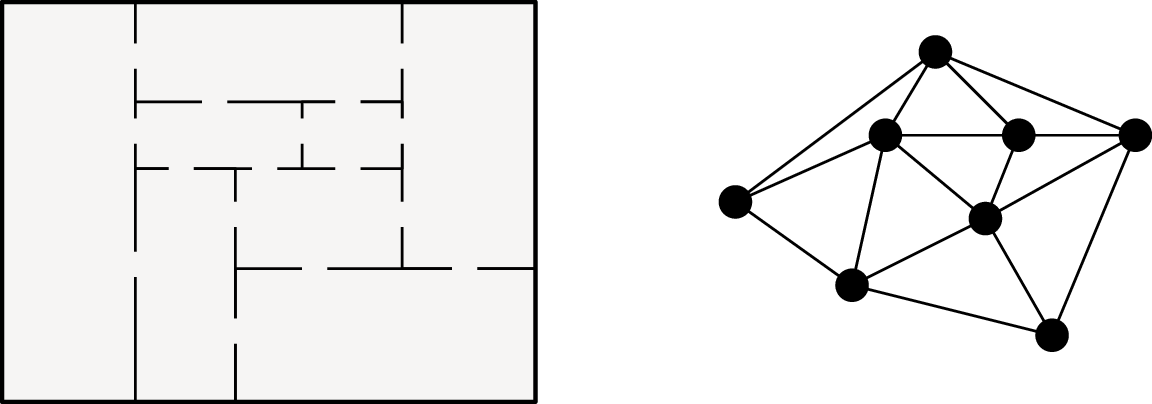}
		\caption{
			Doors between rooms correspond to edges in the weak dual graph.
		}
		\label{fig:rectangulation_doors_and_dual_graph}
	\end{figure}

	The \textsc{Art Gallery Problem}, posed by Klee in 1973 (see O'Rourke~\cite{orourke1987art}), asks for the minimum number of guards placed in a
	polygon required to see its interior.
	Among its many variants, orthogonal polygons
	have received particular attention, motivated in part by applications in settings such as VLSI design, computer graphics, data visualization and architecture.
	Kahn et al.~\cite{kahn1983traditional} showed that every $n$-vertex simple orthogonal polygon can be guarded by $\lfloor n/4 \rfloor$ guards, improving the $\lfloor n/3 \rfloor$ bound for general simple polygons due to Chvátal~\cite{chvatal1975combinatorial}.
	A natural subclass of orthogonal regions consists of \emph{rectangulations} (\Cref{fig:rectangulation_doors_and_dual_graph}).
	
	\begin{definition}[Rectangulation]
		A \emph{rectangulation (rectangular partition)} $\Rcal$ is a finite partition of an axis-aligned rectangular domain $\Omega$ into $n$ closed, non-degenerate, interior-disjoint, axis-aligned rectangles $\Rcal = \{r_1, \dots, r_n\}$, such that no four rectangles meet at a common interior vertex.
	\end{definition}
	
	Guarding rectangulations was initiated by Czyzowicz et al.~\cite{czyzowicz1994guarding}, who considered door-guards placed at shared rectangle boundaries.
	Via a perfect matching argument, they showed that $\lceil n/2 \rceil$ door-guards are always sufficient and occasionally necessary for a rectangulation with $n$ rectangles (see also~\cite{urrutia2000art}).
	Door-guards were similarly considered by Bajuelos et al.~\cite{bajuelos2014guarding}, who extended these bounds to domains allowing non-rectangular environments.
	
	Subsequent work established that several optimization variants are computationally hard.
	Dinitz et al.~\cite{dinitz2007guarding,dinitz2009guarding} proved $\NP$-completeness for vertex-guarding a specified subset of rectangles, as well as for edge-guarding all rectangles with guards restricted to a given subset of edges.
	For the former, they obtained a $4/3$-approximation and proved that any rectangulation admits a polychromatic $3$-coloring (i.e., a vertex $3$-coloring in which every rectangle has all three colors on its boundary).
	They further conjectured an analogous result for four colors, later confirmed by Horev et al.~\cite{horev2009polychromatic} for guillotine rectangulations (i.e., those sliced successively by bisecting lines going from an edge to the opposite) and in general by Dimitrov et al.~\cite{dimitrov2009polychromatic}.
	Dinitz et al.~\cite{dinitz2009guarding} additionally
	proved that, if every rectangle is surrounded by at most five others, including the outer face, then $\lceil n/3 \rceil + 4$ edge-guards suffice.
	On the algorithmic side, Korman et al.~\cite{korman2015line} gave a linear-time
	algorithm for segment-guarding guillotine rectangulations.
	
	These results illustrate a recurring theme in variants of the \textsc{Art Gallery Problem}, namely
	the existence of tight combinatorial bounds while the corresponding optimization problems remain computationally hard.
	Rectangulations thus provide a particularly structured setting in which to study this phenomenon.

	\subsection{Problem Definition}
	
	Even though it is helpful to interpret a rectangulation $\Rcal$ as a
	building $\Omega = \bigcup_{i=1}^n{r_i}$ subdivided into
	rooms connected by doors, our study focuses primarily on the rectangle-vertex incidence.
	Every interior vertex is incident to exactly three rectangles (a \emph{T-junction}), every boundary vertex is incident to at most two,
	and the intersection of any two rectangles is either empty or a non-degenerate line segment (i.e., they meet at exactly two vertices).
	Let $V(\Rcal)$ denote the set of vertices of the partition.
	We assume a standard representation of a rectangulation that gives the vertices of each $r \in \Rcal$ and the set of incident rectangles $C(v) = \{r \in \Rcal : v \in r\}$ of each $v \in V(\Rcal)$.
	This representation has size $\Theta(n)$:
	
	\begin{proposition}
		\label{prop:rectangulation_number_of_vertices}
		A rectangulation $\Rcal$
		with $n$ rectangles has exactly $2n + 2$ vertices.
	\end{proposition}
	
	\begin{proof}
		Consider the graph $G = (V(\Rcal), E)$ formed by the vertices of $\Rcal$, where an edge $\{u,v\} \in E$ exists if $u, v$ are consecutive vertices along a rectangle side.
		Each of the four outer corners of $\Omega$ has degree $2$ in $G$, while all the remaining $t$ vertices have degree $3$.
		The number of vertices is $v = |V(\Rcal)| = 4 + t$, the number of faces (including the outer face) is $f = n+1$ and, by Euler's formula, the number of edges is $e = v + f - 2 = v + n - 1$.
		Since the sum of vertex degrees is $2e = 2 \cdot 4 + 3t$, we have $t = 2n-2$, which implies $v = 2n+2$.
		\qed
	\end{proof}
	
	In the \emph{vertex-guarding} model, a vertex-guard $g \in V(\Rcal)$ covers every rectangle $r \in C(g)$.
	The problem we explore was introduced by Dinitz et al.~\cite{dinitz2007guarding,dinitz2009guarding},
	who called its complete and partial variants \textsc{VGP} and \textsc{PVGP}~\cite{dinitz2007guarding} (or \textsc{GVGP}~\cite{dinitz2009guarding}).
	To unify terminology and distinguish from other variants of the \textsc{Art Gallery Problem}, we refer to them as \textsc{Rectangulation Vertex-Guarding} (\textsc{RVG}) and \textsc{Partial Rectangulation Vertex-Guarding} (\textsc{PRVG}).

	\begin{definition}[\textsc{RVG}]
		Given a rectangulation $\Rcal$ and an integer $k \ge 1$, is there $\Gcal_V \subseteq V(\Rcal)$ with $|\Gcal_V| \le k$ such that every $r \in \Rcal$ is incident to a guard in $\Gcal_V$?
	\end{definition}

	\begin{definition}[\textsc{PRVG}]
		Given a rectangulation $\Rcal$, a target subset of rectangles $\Wcal \subseteq \Rcal$, and an integer $k \ge 1$, is there $\Gcal_V \subseteq V(\Rcal)$ with $|\Gcal_V| \le k$ such that every $r \in \Wcal$ is incident to a guard in $\Gcal_V$?
	\end{definition}

	We use the same notation to refer to both the decision and optimization problems, which ask for the smallest guard set that covers
	$\Rcal$ and $\Wcal$, respectively.

	\subsection{Our Contributions}
	
	Dinitz et al.~\cite{dinitz2007guarding,dinitz2009guarding} proved that \textsc{PRVG} is $\NP$-hard, but left open the complexity status of \textsc{RVG}, asking whether guarding all rectangles might admit a polynomial-time algorithm.
	Their reduction relies on restricting the coverage requirement to $\Wcal \subseteq \Rcal$, which can be thought of as introducing holes to the instance;
	thus, as \textsc{RVG} particularizes \textsc{PRVG} by taking $\Wcal = \Rcal$, the hardness of \textsc{RVG} does not follow immediately from their result.
	They also obtained a $4/3$-approximation via a reduction to $3$-\textsc{Set Cover}, and conjectured whether this factor can be improved.
	Finally, they studied polychromatic colorings of rectangulations with a bounded number of surrounded rectangles.
	
	In this paper, we resolve both questions and address the ``few surrounded rectangles'' case from a guarding perspective.
	Our contributions are as follows:
	
	\begin{itemize}\itemsep0em
		\item \Cref{sec:rectangulations_np_completeness}: We prove that \textsc{RVG} remains $\NP$-complete, even when a fixed subset of vertices is forbidden, by extending their reduction with a hole-filling gadget that replaces holes with constant-sized guillotine partitions.
		
		\item \Cref{sec:approximability_and_ptas}: We design a PTAS for \textsc{RVG} whose running time has a linear dependence on $n$.
		Our technique combines dynamic programming (DP) with the planar shifting technique and rules out $\APX$-hardness unless $\P = \NP$.
		
		\item \Cref{sec:boundary_exposed_rectangulations}: We describe an interesting family of rectangulations in which no rectangle is completely surrounded, prove that these are guillotine partitions and present an $\bigO(n)$-time exact algorithm for \textsc{RVG} that exploits the chordality of their weak duals.
		We generalize this approach to an FPT algorithm for rectangulations with a bounded number of surrounded rectangles.
	\end{itemize}

	\section{NP-Completeness of \textsc{RVG}}
	\label{sec:rectangulations_np_completeness}
	
	Polynomial-time solvable geometric covering problems often become intractable~in the presence of holes (e.g.,~\cite{biedl2017guarding,chin1986optimum,cruz2022algorithms,cruz2022on,fekete2025guarding}), but we show that holes are not required for vertex-guarding rectangulations to become hard.
	Specifically, we show:
	\begin{theorem}
		\label{thm:rectangulation_np_completeness}
		\textsc{RVG} is $\NP$-complete.
	\end{theorem}
	
	Membership in $\NP$ is immediate: given $\Rcal$ and $\Gcal$, one can verify in~$\bigO(|\Rcal|)$ time that $|\Gcal| \le k$ and every $r \in \Rcal$ is incident to a guard in $\Gcal$.
	To show $\NP$-hardness, we build upon the polynomial-time reduction of Dinitz et al.~\cite{dinitz2007guarding,dinitz2009guarding}, which maps a special instance~$G$ of \textsc{Minimum Vertex Cover}
	to a rectangulation $\Rcal_G$ together with a target subset $\Wcal_G$ of rectangles that must be covered.

	Our goal is to transition from \textsc{PRVG} to \textsc{RVG} while preserving the combinatorial structure of the reduction.
	The main challenge is to ensure that each hole is guarded independently, preventing
	guards inside holes from helping cover rectangles in $\Wcal_G$ and vice versa.
	To achieve this isolation, we design a gadget $\Phi$ (\Cref{fig:rectangulation_np_hardness_hole_gadget}) and replace each hole $h$
	by a suitably perturbed copy $\Phi_h$ of $\Phi$.

	\subsection{Original Reduction for \textsc{PRVG}}
	
	For completeness, we briefly recall the reduction of Dinitz et al.~\cite{dinitz2007guarding,dinitz2009guarding} on which our construction builds.
	Given a non-bipartite, $3$-connected, cubic planar graph $G$ of girth greater than $3$, they construct its line graph $L(G)$,
	triangulate every non-triangular face to produce a proper plane triangulation $L_1(G)$, and then construct in linear time a rectangular dual $\Rcal_G$ of $L_1(G)$ using a result of Kozminski and Kinnen~\cite{kozminski1984an}.
	The rectangles are partitioned into \emph{non-hole} rectangles $\Wcal_G$, bijectively corresponding to the edges of $G$, and \emph{hole} rectangles $\Hcal_G = \Rcal_G \setminus \Wcal_G$, bijectively mapping to the internal faces of $G$.
	Moreover, no two hole rectangles touch, and the vertices of $G$ are in one-to-one correspondence with the vertices of $\Rcal_G$ incident to exactly three non-hole rectangles and zero hole rectangles.
	Thus, placing a vertex-guard at such a $(3,0)$-meeting point covers the three non-hole rectangles corresponding to the edges incident to the associated vertex in $G$, and is equivalent to selecting that vertex in the \textsc{Minimum Vertex Cover} instance.

	\subsection{Reduction for \textsc{RVG}}
	
\begin{figure}[hbpt]
		\centering
		\includegraphics[scale=0.6]{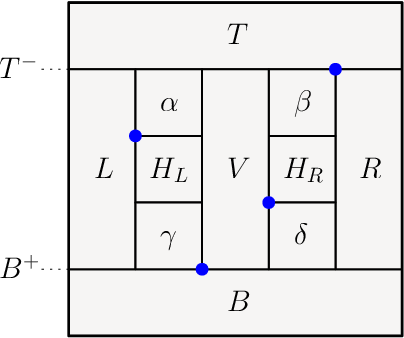}
		\caption{Hole gadget $\Phi$ with an optimal set of four local guards.}
		\label{fig:rectangulation_np_hardness_hole_gadget}
	\end{figure}

	Without loss of generality, consider the normalized square $[0, 5] \times [0, 5]$, representing a hole.
	The gadget $\Phi$ replacing this hole consists of $11$ rectangles (\Cref{fig:rectangulation_np_hardness_hole_gadget}):
	\begin{itemize}\itemsep0em
		\item \emph{Top and bottom shells:} $T = [0,5] \times [4,5]$ and $B = [0,5] \times [0,1]$.
		\item \emph{Left and right shells:} $L = [0,1] \times [1,4]$ and $R = [4,5] \times [1,4]$.
		\item \emph{Core:} $\alpha = [1,2] \times [3,4]$, $\beta = [3,4] \times [3,4], \gamma = [1,2] \times [1,2]$ and $\delta = [3,4] \times [1,2]$.
		\item \emph{Separators:} $H_L = [1,2] \times [2,3], H_R = [3,4] \times [2,3]$ and $V = [2,3] \times [1,4]$.
	\end{itemize}
	
	Since every interior vertex of $\Phi$ is a T-junction, $\Phi$ is a valid rectangulation.
	The following lemma establishes a tight bound on the number of guards for $\Phi$.

	\begin{lemma}
		\label{lemma:rectangulation_np_hardness_hole_gadget_4_local_guards}
		Exactly $4$ local guards are necessary and sufficient to cover $\Phi$.
	\end{lemma}

	\begin{proof}
		For necessity, since the four core rectangles $\alpha, \beta, \gamma, \delta$ are pairwise vertex-disjoint, a single vertex-guard can cover at most one of them.
		Moreover, every core rectangle is separated from the boundary $\partial \Phi$ by shell rectangles, so no guard placed on $\partial \Phi$ is incident to any core rectangle.
		Thus, at least four guards must be located inside $\Phi$.
		For sufficiency, every rectangle of $\Phi$ is incident to at least one of the four vertices shared by $(\alpha, L, H_L)$, $(\beta, T, R)$, $(\gamma, V, B)$ and $(\delta, V, H_R)$.
		\qed
	\end{proof}
	
	Note that, when embedding $\Phi$ into a hole $h \in \Hcal_G$, T-junctions of the surrounding rectangulation may already occur on $\partial h$, so we have to prevent them
	from creating four-way intersections with internal cuts of $\Phi$.
	The only internal cuts that can originate such intersections are the bottom edge $T^-$ of $T$ and the top edge $B^+$ of $B$, since all other cuts are contained in $\inter \Phi$.
	There are only finitely many $y$-coordinates of vertices on $\bd h$ that must be avoided, and every pair of holes in $\Hcal_G$ is disjoint by construction.
	Hence, we can independently choose two
	sufficiently small offsets for $T^-$ and $B^+$ such that the perturbed cuts
	preserve their relative order with respect to other horizontal cuts.
	
	\begin{lemma}
		There is a combinatorially equivalent copy $\Phi_h$ of $\Phi$, with $\bd \Phi_h = \bd h$, such that replacing each $h \in \Hcal_G$ with $\Phi_h$ in $\Rcal_G$ yields a valid rectangulation.
	\end{lemma}

	Let $\Rcal'$ denote the rectangulation obtained by replacing every hole $h \in \Hcal_G$.
	The reduction remains polynomial as $\Phi_h$ is a constant-size guillotine partition, so we conclude by proving its correctness, from which $\NP$-hardness follows.
	
	\begin{theorem}
		The \textsc{PRVG} instance $\Rcal_G = \Wcal_G \cup \Hcal_G$ admits a vertex-guard set of size at most $k$ for $\Wcal_G$ if and only if the \textsc{RVG} instance $\Rcal'$ admits a vertex-guard set of size at most $k + 4|\Hcal_G|$.
	\end{theorem}
	
	\begin{proof}
		$(\Rightarrow)$
		Let $\Gcal$ be a vertex-guard set of size at most $k$ covering $\Wcal_G$.
		According to~\Cref{lemma:rectangulation_np_hardness_hole_gadget_4_local_guards}, we extend $\Gcal$ by placing four local guards inside each gadget $\Phi_h$, thereby covering $\Hcal_G$.
		The new guard set $\Gcal'$ is therefore a valid solution for $\Rcal'$ of size $|\Gcal| + 4|\Hcal_G|$.
		$(\Leftarrow)$
		Conversely, let $\Gcal'$ be a solution for $\Rcal'$ of size at most $k + 4|\Hcal_G|$.
		Each gadget $\Phi_h$ contains four pairwise vertex-disjoint core rectangles, none of which can be covered by a guard on $\bd \Phi_h$ or outside $\Phi_h$, as per~\Cref{lemma:rectangulation_np_hardness_hole_gadget_4_local_guards}.
		Thus, $\Gcal'$ must contain at least four local guards strictly inside each gadget, not incident to any rectangle in $\Wcal_G$.
		Removing all these strictly interior guards leaves a set $\Gcal \subseteq \Gcal'$ of at most $|\Gcal'| - 4|\Hcal_G| \le k$ guards that still covers $\Wcal_G$.
		\qed
	\end{proof}

	Dinitz et al.~\cite{dinitz2007guarding,dinitz2009guarding} also proved hardness for edge-guarding complete rectangulations when a fixed subset of edges is forbidden.
	In the spirit of this result, our construction similarly extends to a constrained variant of \textsc{RVG} in which not every vertex is allowed to hold a guard: by forbidding any fixed set of vertices in $\Phi$ that are unused by the solution of~\Cref{lemma:rectangulation_np_hardness_hole_gadget_4_local_guards}, the same reduction holds.

	\begin{theorem}
		\textsc{RVG} is $\NP$-complete even with a fixed forbidden subset of vertices.
	\end{theorem}

	\section{Approximability and PTAS}
	\label{sec:approximability_and_ptas}
	
	Being hard to solve exactly, it is natural to ask whether \textsc{RVG}
	also remains inapproximable.
	As it turns out, the problem has enough structure to admit constant-factor approximations: since each vertex is incident to at most three rectangles, \textsc{RVG} is a special case of $3$-\textsc{Set Cover}.
	The standard greedy algorithm already yields an $11/6$-approximation~\cite{johnson1973approximation},
	which can be improved to $4/3$ by the semi-local optimization algorithm of Duh and Fürer~\cite{duh1997approximation}.
	Dinitz et al.~\cite{dinitz2009guarding} questioned whether this factor could still be tightened by exploiting geometry.
	
	We answer this
	positively
	by designing a PTAS for \textsc{RVG}.
	Our approach relies on planarity, which is lost
	in generic \textsc{Set Cover}.
	Rather than addressing \textsc{RVG} directly, we express it as a
	\textsc{Red-Blue Dominating Set} problem on planar graphs (\textsc{PRBDS}),
	strengthening our result to a broader graph-theoretic setting.

	\begin{definition}[\textsc{Planar Red-Blue Dominating Set}]%
		Given a planar bipartite graph $G = (R \cup B, E)$ and an integer $k \ge 1$, is there $S \subseteq B$ with $|S| \le k$ such that every $r \in R$ has a neighbor $b \in S$?
	\end{definition}
	
	In the following subsections, we develop a PTAS for \textsc{PRBDS} using Baker's shifting technique~\cite{baker1994approximation} together with dynamic programming over tree decompositions of bounded-outerplanarity graphs.
	Baker's technique is based on partitioning the vertices of a planar graph into breadth-first search (BFS) layers from a source vertex.
	For a fixed integer $k$, it considers $k$ different shifts, each dividing the graph into slices of at most $k$ consecutive layers.
	For each shift, the optimization problem is solved optimally on every slice, taking into account the adjacent layers to preserve interactions across slice boundaries, and the solutions for the individual slices are then combined into a solution for the original graph.
	Selecting the best solution across all $k$ shifts guarantees the desired approximation ratio and, because the resulting subproblems have bounded width (depending on $k$) and are solvable in polynomial time, choosing $k$ appropriately yields a PTAS.
	
	The \emph{(rectangle-vertex) incidence graph} $H(\Rcal) = (V_H, E_H)$, where $V_H = \Rcal \cup V(\Rcal)$ and $\{r,v\} \in E_H \Leftrightarrow v \in r$, plays a relevant role in our analysis.
	Since $|V(\Rcal)| = 2n+2$,
	$|V_H| \in \Theta(n)$.
	Placing each vertex $v \in V(\Rcal)$ at its geometric coordinates and each $r \in \Rcal$ at the center of its rectangle yields a planar embedding of $H(\Rcal)$.
	As $H(\Rcal)$ is planar bipartite,
	\textsc{RVG} is exactly \textsc{PRBDS} on
	$H(\Rcal)$
	when $R = \Rcal$ and $B = V(\Rcal)$,
	so it follows from~\Cref{thm:rectangulation_np_completeness} that \textsc{PRBDS} is $\NP$-complete.
	The PTAS applies to \textsc{PRVG} as well, by setting $R = \Wcal \subseteq \Rcal$.

	\begin{theorem}
		For any $\varepsilon > 0$, a $(1+\varepsilon)$-approximation for an $n$-vertex instance of \textsc{PRBDS} can be computed in $2^{\bigO(1/\varepsilon)} \cdot n$ time.
	\end{theorem}

	\subsection{Shifting Algorithm and Approximation Ratio}
	
\begin{figure}[tb]
		\centering
		\includegraphics[scale=0.75]{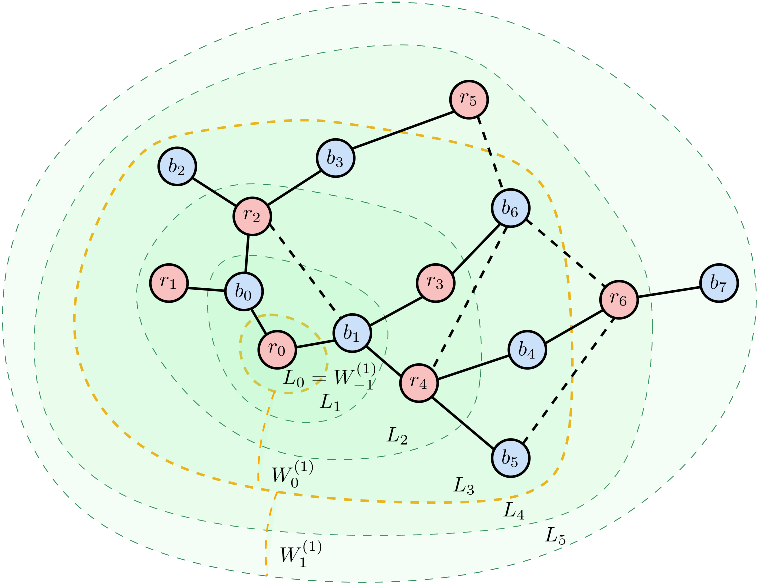}
		\caption{
			BFS layers $L_0, \dots, L_5$ from source $r_0 \in R$ partitioned into target blocks $W_m^{(i)}$ of size $k = \lceil 2/\varepsilon \rceil = 3$, for shift index $i = 1$.
			Each component $C_m^{(1)}$ extends its target block $W_m^{(1)}$ with two adjacent boundary layers (e.g., $C_0^{(1)} = W_0^{(1)} \cup L_0 \cup L_4$).
		}
		\label{fig:planar_red_blue_dominating_set_ptas_shifting_components}
	\end{figure}

	Let $G = (R \cup B, E)$ be a connected planar bipartite graph with red ($R$) and blue ($B$) vertices.
	We partition $R \cup B$ into BFS layers from any source $s \in R \cup B$.
	Let $L_j$ be the set of vertices at distance $j$ from~$s$ for $j \ge 0$, and let $L_j = \emptyset$ for $j < 0$.
	Given an approximation error $\varepsilon > 0$, set $k = \lceil 2/\varepsilon \rceil$.
	For each fixed shift index $0 \le i < k$, we partition the layers into disjoint \emph{target blocks} $W_m^{(i)}$ of $k$ consecutive layers (the first and last blocks may exceptionally contain fewer).
	Formally, define the $m$-th target block, for $m \ge 0$, as $W_m^{(i)} = \bigcup_{j = mk+i}^{(m+1)k+i-1}{L_j}$, and, when $i > 0$, additionally define $W_{-1}^{(i)} = \bigcup_{j=0}^{i-1}{L_j}$,
	with $i$ layers
	(\Cref{fig:planar_red_blue_dominating_set_ptas_shifting_components}).

	BFS guarantees that $v \in L_{j-1} \cup L_j \cup L_{j+1}$ for any edge $\{u,v\} \in E$ with $u \in L_j$.
	Hence, any $r \in R \cap W_m^{(i)}$ can only be dominated by a blue vertex in $W_m^{(i)}$ or in adjacent layers.
	In this sense, we define the \emph{component} $C_m^{(i)} = W_m^{(i)} \cup L_{mk+i-1} \cup L_{(m+1)k+i}$ and, for $i > 0$, $C_{-1}^{(i)} = W_{-1}^{(i)} \cup L_i$,
	which spans at most $k+2$ consecutive BFS layers of a planar graph.
	Therefore, the induced subgraph $G[C_m^{(i)}]$ is $(k+3)$-outerplanar and has bounded treewidth $\tw(G[C_m^{(i)}])$~\cite{bodlaender1998partial}. %
	\begin{proposition}
		\label{prop:induced_component_treewidth}
		$\tw(G[C_m^{(i)}]) \in \bigO(k)$.
	\end{proposition}
	
	As we detail in~\Cref{sec:ptas_treewidth_dynamic_programming}, the next step is to compute a nice tree decomposition
	of width $\bigO(k)$ for each induced subgraph $G[C_m^{(i)}]$.
	Using dynamic programming, we compute a local solution $S_m^{(i)} \subseteq B \cap C_m^{(i)}$ of minimum cardinality that dominates all vertices in $R \cap W_m^{(i)}$.
	Red vertices located in
	$L_{mk+i-1} \cup L_{(m+1)k+i}$ are not required to be dominated by $S_m^{(i)}$, as they belong to adjacent target blocks and will already be dominated by their respective local solutions.
	However, blue vertices in these boundary layers may be included in $S_m^{(i)}$ to dominate vertices within $R \cap W_m^{(i)}$.
	The global solution $S^{(i)} = \bigcup_{m \ge -1}{S_m^{(i)}}$ for a fixed shift~$i$ is obtained by taking the union of all the local solutions.
	Because target blocks
	partition $R \cup B$ for any fixed shift $i$, every $r \in R$ belongs to exactly one target block $W_m^{(i)}$ and is dominated by the corresponding local solution $S_m^{(i)}$.
	This ensures that every
	$S^{(i)}$ is a feasible red-blue dominating set for $G$.
	Repeating for all $k$ shifts, the algorithm returns the smallest
	global solution found.

	\begin{lemma}
		The smallest solution found is a $(1 + \varepsilon)$-approximation for \textsc{PRBDS}.
	\end{lemma}
	
	\begin{proof}
		Let $S^\star \subseteq B$ be an optimal red-blue dominating set for $G$.
		For a fixed shift~$i$ and index $m$,
		$C_m^{(i)}$ contains every neighbor of any $r \in R \cap W_m^{(i)}$, since every edge joins vertices in the same or adjacent layers.
		Consequently, the subset $S^\star \cap C_m^{(i)}$ is guaranteed to dominate all vertices in $R \cap W_m^{(i)}$.
		By hypothesis, our dynamic programming finds a minimum-cardinality local solution $S_m^{(i)} \subseteq B \cap C_m^{(i)}$ that dominates $R \cap W_m^{(i)}$, hence $|S_m^{(i)}| \le |S^\star \cap C_m^{(i)}|$.
		Two consecutive components $C_{m-1}^{(i)}$ and $C_m^{(i)}$ overlap on at most two layers: $L_{mk+i-1}$ and $L_{mk+i}$.
		Let $F^{(i)} = \bigcup_{m \ge 0} \left( L_{mk+i-1} \cup L_{mk+i} \right)$ be the union of these
		boundary layers for a given shift~$i$.
		Since every vertex outside $F^{(i)}$ belongs to exactly one component and those in $F^{(i)}$ belong to at most two,
		the global solution $S^{(i)} = \bigcup_m S_m^{(i)}$ for a fixed shift $i$ has total size $|S^{(i)}| \le \sum_m |S_m^{(i)}| \le \sum_m |S^\star \cap C_m^{(i)}| \le |S^\star| + |S^\star \cap F^{(i)}|$.
		As~$i$ ranges over $0, \dots, k-1$, each layer is assigned to an overlap~$F^{(i)}$ at most twice.
		Thus, $\sum_{i=0}^{k-1} |S^\star \cap F^{(i)}| \le 2 |S^\star|$ and,
		by the pigeonhole principle, for some shift~$i^\star$ the overlap is bounded by the average:
		$|S^\star \cap F^{(i^\star)}| \le \frac{2 |S^\star|}{k}$.
		Because $2/k \le \varepsilon$, the total size for this shift is bounded by $|S^{(i^\star)}| \le \left(1+\frac{2}{k}\right)|S^\star| \le (1 + \varepsilon) |S^\star|$.
		\qed
	\end{proof}

	\subsection{Treewidth-Based Dynamic Programming}
	\label{sec:ptas_treewidth_dynamic_programming}
	
	We now describe our DP algorithm for computing a local optimal solution for each component $C_m^{(i)}$.
	As the algorithm applies to a fixed pair $(m,i)$, we use~$C$ and $W$ instead of $C_m^{(i)}$ and $W_m^{(i)}$ in the sequel.
	By~\Cref{prop:induced_component_treewidth}, $G[C]$ has
	treewidth $t \in \bigO(k)$.
	We start by building a tree decomposition of width $\bigO(t)$ in $2^{\bigO(t)} \cdot |C|$ time~\cite{bodlaender2016a},
	and then convert it into a \emph{nice tree decomposition} $(T, \Xcal)$ of the same width that contains $\bigO(t \cdot |C|)$ nodes, in $\bigO(t^2 \cdot |C|)$ time~\cite{cygan2015parameterized,kloks1994treewidth}.

	As usual, $\tw(G)$ stands for the treewidth of a graph $G = (V, E)$, and a tree decomposition $(T, \Xcal)$ is defined by the following axioms:
	\emph{(1)} $\bigcup_{i \in V(T)}{X_i} = V$;
	\emph{(2)} for every $\{u,w\} \in E$, there exists a node $i \in V(T)$ such that $u, w \in X_i$;
	\emph{(3)} for every $v \in V$,
	$T_v = \{i \in V(T) : v \in X_i\}$ induces a connected subtree of~$T$.
	A rooted tree decomposition $(T, \Xcal)$ is \emph{nice} if $X_{\textsf{root}} = \emptyset$, $X_\ell = \emptyset$ for every leaf $\ell \in V(T)$, and every non-leaf node $i \in V(T)$ is of one of three types:
	\emph{(1)} an \emph{introduce node}, having a single child $j$ with $X_i = X_j \cup \{v\}$ for some $v \notin X_j$;
	\emph{(2)} a \emph{forget node}, having a single child $j$ with $X_i = X_j \setminus \{v\}$ for some $v \in X_j$; or
	\emph{(3)} a \emph{join node}, having two children $l$ and $r$ with $X_i = X_l = X_r$.
	Here, we use $X_i \subseteq C$ to denote the bag of a node $i \in V(T)$ and $V_i$ to denote the union of the bags in the subtree rooted at $i$.
	We assume that the \textsf{root} node representing the whole $G[C]$ is a forget node with an empty bag (e.g., by adding a sequence of forget nodes that empty the original root while preserving width).
	
	The idea of the algorithm is to process $T$ bottom-up, maintaining a table of valid configurations for each node.
	A \emph{configuration} $c$ for a node $i \in V(T)$ assigns a state to each vertex $v \in X_i$.
	A blue vertex $v \in X_i$
	is assigned a state $c(v) \in \{\textsf{IN}, \textsf{OUT}\}$ indicating whether it is selected, while a red vertex $v \in X_i \cap W$
	is assigned a state $c(v) \in \{\textsf{DOM}, \textsf{UNDOM}\}$ according to whether it is already dominated by a selected blue vertex in $V_i$.
	Red boundary vertices in $X_i \setminus W$ require no domination and thus are assigned the fixed state $\textsf{IGN}$ (ignored).

	We keep a table $DP[i,c]$ storing the minimum size of a blue subset $S_i \subseteq B \cap V_i$ consistent with the configuration $c$, meaning that:
	\emph{(1)} for all $v \in B \cap X_i$, $v \in S_i$ iff $c(v) = \textsf{IN}$;
	\emph{(2)} for all $v \in R \cap V_i \cap W$, if $v \notin X_i$ or $c(v) = \textsf{DOM}$, then $v$ is adjacent to some $u \in S_i$.
	If no such $S_i$ exists, we set $DP[i,c] = \infty$.
	The minimum for $C$ is given by $DP[\textsf{root}, \emptyset]$,
	being $DP[i,c]$ computed recursively
	as follows.

	\begin{itemize}
		\item \emph{Leaf node.}
		If $i$ is a leaf, $X_i = \emptyset$ and we set $DP[i, \emptyset] = 0$.
		
		\item \emph{Introduce node.}
		Let $i$ be an introduce node with child $j$, where $X_i = X_j \cup \{v\}$, and let $N(v)$ be the open neighborhood of $v$.
		For any configuration $c$ of $X_i$:
		\begin{itemize}\itemsep0em
			\item If $v \in B$ and $c(v) = \textsf{OUT}$, the introduction of $v$ preserves the domination status of every red vertex, so $DP[i,c] = DP[j, c \setminus \{(v, \textsf{OUT})\}]$.
			\item If $v \in B$ and $c(v) = \textsf{IN}$, any red neighbor $u \in N(v) \cap R \cap W \cap X_j$ with $c(u) = \textsf{UNDOM}$ contradicts the selection of $v$ and makes $DP[i,c] = \infty$.
			Otherwise, $DP[i,c] = \min_{c'}(DP[j, c'] + 1)$, with the minimum taken over all configurations $c'$ of $X_j$ compatible with $c$: $c'(x) = c(x)$ for all $x \in X_j \setminus N(v)$, and $c'(u) \in \{\textsf{DOM}, \textsf{UNDOM}\}$ for all $u \in N(v) \cap R \cap W \cap X_j$.
			\item If $v \in R \cap W$,
			as $v$ has no neighbor in $V_j \setminus X_j$, any blue vertex in $V_i$ dominating $v$ lies in $X_i$.
			If there exists $u \in N(v) \cap X_i$ with $c(u) = \textsf{IN}$, $c(v)$ must be $\textsf{DOM}$, otherwise $\textsf{UNDOM}$.
			Configurations satisfying such conditions map to $DP[i,c] = DP[j, c \setminus \{(v, c(v))\}]$, and invalid ones to~$\infty$.
			\item If $v \in R \setminus W$, i.e., $v$ is a boundary red vertex, we enforce $c(v) = \textsf{IGN}$ and set $DP[i,c] = DP[j, c \setminus \{(v, \textsf{IGN})\}]$.
		\end{itemize}
		
		\item \emph{Forget node.}
		Let $i$ be a forget node with child $j$, where $X_i = X_j \setminus \{v\}$.
		The configuration $c$ on $X_i$ defines the states of all vertices in $X_j$ except for $v$.
		\begin{itemize}\itemsep0em
			\item If $v \in R \cap W$, $v$ will never appear again in any ancestor bag, so it must already be dominated at this step, as there will be no future opportunity to do so: $DP[i,c] = DP[j, c \cup \{(v, \textsf{DOM})\}]$.
			\item If $v \in R \setminus W$, we set $DP[i,c] = DP[j, c \cup \{(v, \textsf{IGN})\}]$.
			\item If $v \in B$, we minimize over two possibilities (include $v$ or exclude $v$ from the solution): $DP[i,c] = \min(DP[j, c \cup \{(v, \textsf{IN})\}], DP[j, c \cup \{(v, \textsf{OUT})\}])$.
		\end{itemize}

		\item \emph{Join node.}
		Let $i$ be a join node with children $l$ and $r$, where $X_i = X_l = X_r$.
		We merge solutions from the two subtrees by subtracting their blue vertex overlap.
		As such, for a given configuration $c$ of $X_i$, $DP[i,c]$ is set to $\min_{c_l, c_r}(DP[l, c_l] + DP[r, c_r] - |\{u \in B \cap X_i : c(u) = \textsf{IN}\}|)$.
		The minimum is evaluated over
		pairs $(c_l, c_r)$ for $X_l$ and $X_r$ consistent with~$c$, meaning that $c_l(u) = c_r(u) = c(u)$ for all $u \in B \cap X_i$, and $c_l(u) = c_r(u) = \textsf{IGN}$ for all $u \in R \cap X_i \setminus W$.
		For red target vertices $u \in R \cap X_i \cap W$, if $c(u) = \textsf{UNDOM}$, then neither subtree can provide a dominating vertex for $u$ and $c_l(u) = c_r(u) = \textsf{UNDOM}$; otherwise, $c(u) = \textsf{DOM}$ and then $u$ must be dominated by a blue vertex selected in at least one subtree, which restricts $(c_l(u), c_r(u))$ to $\{(\textsf{DOM}, \textsf{UNDOM}), (\textsf{UNDOM}, \textsf{DOM}), (\textsf{DOM}, \textsf{DOM})\}$.
	\end{itemize}

	\paragraph{Complexity analysis.}
	The nice tree decomposition $T$ has $\bigO(t \cdot |C|)$ nodes.
	Each vertex has at most two valid states,
	so each node stores
	$2^{\bigO(t)}$ dynamic programming entries, requiring space $\bigO(2^{\bigO(t)} \cdot t \cdot |C|) = \bigO(2^{\bigO(k)} \cdot |C|)$.
	Processing leaf, forget, red-introduce, and blue-introduce (with $c(v) = \textsf{OUT}$) nodes takes $\bigO(t)$ time per configuration, hence $2^{\bigO(t)}$ time per node.
	For a blue-introduce node with $c(v) = \textsf{IN}$,
	the transition checks
	two compatible prior states (\textsf{DOM} and \textsf{UNDOM}) for each red target neighbor $u$ with $c(u) = \textsf{DOM}$.
	Summing over $d$ red targets marked \textsf{DOM} gives at most $\sum_{d=0}^{|X_i|}{\binom{|X_i|}{d} 2^d 1^{|X_i|-d}} = 3^{|X_i|} \in 2^{\bigO(k)}$ transition evaluations.
	For a join node, merging the states of target vertices in two subtrees requires evaluating three valid predecessor state pairs $(c_l(u), c_r(u))$ for every $u$ with $c(u) = \textsf{DOM}$, and one
	for each $u$ with $c(u) = \textsf{UNDOM}$.
	Summing over all valid bag configurations yields $\sum_{d=0}^{|X_i|}{\binom{|X_i|}{d} 3^d 1^{|X_i|-d}} \in 2^{\bigO(k)}$ evaluations.
	Therefore, the total time to compute an optimal solution for one $G[C]$ is $\bigO(k^2 \cdot |C| + 2^{\bigO(k)} \cdot k \cdot |C|) = \bigO(2^{\bigO(k)} \cdot |C|)$.
	Because
	every vertex belongs to at most two components,
	$\sum_m{|C_m|} \in \bigO(|V(G)|)$.
	Repeating over the $k = \lceil 2/\varepsilon \rceil$ shifts
	gives the bound $\bigO\left(\sum_{i=0}^{k-1}{\left(2^{\bigO(k)} \sum_m{|C_m^{(i)}|}\right)}\right) = \bigO(k 2^{\bigO(k)} \bigO(|V(G)|)) = 2^{\bigO(1/\varepsilon)} \cdot n$ for the PTAS.

	\section{Boundary-Exposed Rectangulations}
	\label{sec:boundary_exposed_rectangulations}
	
\begin{figure}[ht]
		\centering
		\begin{subfigure}[t]{0.45\textwidth}
			\centering
			\includegraphics[scale=0.5]{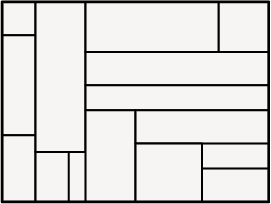}
			\caption{}
			\label{fig:boundary_exposed_rectangulation}
		\end{subfigure}
		\begin{subfigure}[t]{0.45\textwidth}
			\centering
			\includegraphics[scale=0.5]{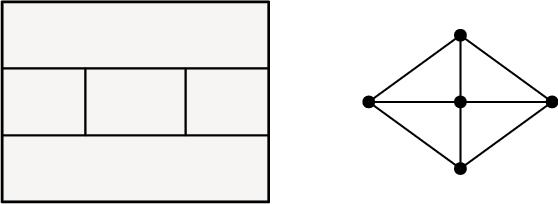}
			\caption{}
			\label{fig:guillotine_partition_not_chordal_counterexample}
		\end{subfigure}
		\caption{
			(a) A boundary-exposed rectangulation.
			(b) Guillotine rectangulation whose weak dual contains an induced $4$-cycle.
		}
		\label{fig:boundary_exposed_rectangulation_and_chordal_counterexample}
	\end{figure}

	In this section, we identify a natural subclass of rectangulations that can be solved efficiently.
	They
	can be thought of as floor plans where every room has a window to the exterior (\Cref{fig:boundary_exposed_rectangulation}).
	We make use of the \emph{(weak) dual} graph $D(\Rcal) = (V_D, E_D)$, which has a vertex $v_i$ for each $r_i \in \Rcal$ and an edge $\{v_i, v_j\}$ when
	$\{r_i, r_j\} \subseteq C(v)$ for some $v \in V(\Rcal)$ (see~\Cref{fig:rectangulation_doors_and_dual_graph}).
	
	\begin{definition}[Boundary-exposed rectangulation]
		A rectangulation $\Rcal$ of $\Omega$ is \emph{boundary-exposed}
		if every rectangle $r \in \Rcal$ has a side contained in $\bd\Omega$.
	\end{definition}
	
	In~\Cref{sec:boundary_exposed_rectangulations_linear_time}
	we prove that
	every boundary-exposed rectangulation $\Rcal$ is guillotine and admits a chordal outerplanar weak dual $D(\Rcal)$.
	These properties reduce guarding to \textsc{Minimum Clique Cover} on a chordal graph.
	We then extend this tractability to a broader range of instances in~\Cref{sec:boundary_exposed_rectangulations_generalization}.

	\subsection{Linear-Time Algorithm for Boundary-Exposed Rectangulations}
	\label{sec:boundary_exposed_rectangulations_linear_time}
	
	We first uncover the guillotine structure of boundary-exposed rectangulations;
	not every guillotine partition has a chordal dual, though (see~\Cref{fig:guillotine_partition_not_chordal_counterexample}).
	
	\begin{lemma}
		\label{lemma:boundary_exposed_rectangulation_contains_guillotine_cut}
		Let $\Rcal$ be a boundary-exposed rectangulation of $\Omega$ with $n > 1$ rectangles.
		Then $\Rcal$ contains a
		guillotine cut.
	\end{lemma}
	
	\begin{proof}
		Let $\Gamma = \left(\bigcup_{r \in \Rcal}{\partial r}\right) \setminus \partial\Omega$ be the union of internal sides, and let~$\Lcal$ be the set of maximal axis-aligned
		line segments contained in $\Gamma$, extended as far as possible without changing direction or entering a rectangle.
		Since $n > 1$, $\Lcal \ne \emptyset$.
		Assume for
		contradiction that $\Rcal$ has no guillotine cut, so every maximal segment $\ell \in \Lcal$ has at least one endpoint $t \in \inter \Omega$.
		As no four rectangles meet,
		$t$ is a T-junction where $\ell$ meets a perpendicular
		$\ell' \in \Lcal$ such that $t \in \ell \cap \inter \ell'$.
		Build a directed graph $G = (\Lcal, E)$ with an
		edge $(\ell, \ell') \in E$ whenever an endpoint of $\ell$ lies in $\inter \ell'$.
		Under the assumption that no guillotine cut exists, every vertex in $G$ has out-degree at least $1$.
		Given that $\Lcal$ is finite, $G$
		contains a directed cycle $(\ell_1, \ell_2, \dots, \ell_m, \ell_1)$ which forms a closed orthogonal chain $\gamma \subset \inter \Omega$.
		Since $\Gamma \cap \partial\Omega = \emptyset$,
		$\gamma$ bounds a non-empty region $U \subset \inter \Omega$ containing at least one rectangle, which is impossible.
		Therefore a guillotine cut must exist.
		\qed
	\end{proof}
	
	\begin{theorem}
		\label{thm:boundary_exposed_rectangulation_guillotine}
		Every boundary-exposed rectangulation is a guillotine partition.
	\end{theorem}

	\begin{proof}
		We proceed by strong induction on
		$n = |\Rcal|$.
		For $n = 1$, $\Rcal = \{\Omega\}$ is trivially guillotine.
		By~\Cref{lemma:boundary_exposed_rectangulation_contains_guillotine_cut}, a guillotine cut $\ell^\star$ partitions $\Omega$ into
		rectangular subdomains $\Omega_1, \Omega_2$ and $\Rcal$ into
		sub-rectangulations $\Rcal_1, \Rcal_2$ with $|\Rcal_1|, |\Rcal_2| < n$.
		Since $\Rcal$ is boundary-exposed, every $r \in \Rcal_i$, with $i \in \{1,2\}$, shares a non-degenerate
		segment $\ell$ with $\partial \Omega$.
		Because $r \subseteq \Omega_i$, we have $\ell \subseteq \partial\Omega_i$, where $\partial\Omega_i = (\partial\Omega \cap \Omega_i) \cup \ell^\star$.
		Thus $r$ shares a boundary with $\partial\Omega_i$, meaning both $\Rcal_1$ and $\Rcal_2$ are boundary-exposed rectangulations of $\Omega_1$ and $\Omega_2$, respectively.
		By induction, both
		are guillotine partitions separated by a cut $\ell^\star$, therefore so is $\Rcal$.
		\qed
	\end{proof}

	\begin{lemma}
		\label{lemma:boundary_exposed_rectangulation_dual_graph_chordal_outerplanar}
		For every boundary-exposed rectangulation $\Rcal$, its weak dual $D(\Rcal)$ is outerplanar and chordal.
	\end{lemma}
	
	\begin{proof}
		That $D(\Rcal)$ is outerplanar follows from the existence of an outer walk that contains every vertex by traversing $\bd \Omega$ counter-clockwise.
		Alternatively, construct a planar embedding for $D(\Rcal)$ with all vertices on $\bd\Omega$ as follows.
		Place a vertex $v_\infty$ outside $\Omega$ representing the outer face, and, for each $r_i \in \Rcal$, place its
		dual vertex $v_i$ at its center.
		Since $r_i \cap \bd\Omega$ contains a non-degenerate line segment~$\ell$, let $p_i$ be the midpoint of any such $\ell$ and draw a curve connecting~$v_i$ to $v_\infty$ through $p_i$.
		The curve portions inside $\Omega$ are pairwise disjoint, as the rectangles have disjoint interiors, while
		the outer portions $p_i v_\infty$ can be drawn non-crossing by ordering the $p_i$ along the convex boundary $\bd \Omega$.
		Thus $D(\Rcal) \cup \{v_\infty\}$ is planar with every $v \in V_D$ incident to $v_\infty$, so $D(\Rcal)$ is outerplanar.
		
		Now suppose for contradiction that $D(\Rcal)$ contains an induced cycle $C$ of length $k \ge 4$.
		In an outerplanar embedding of $D(\Rcal)$, $C$ bounds a region containing no dual vertex
		$v_i \in V_D$ (otherwise any such vertex would correspond to a rectangle not touching $\bd \Omega$), which in fact is a bounded face with at least four sides.
		However, since at most three rectangles meet at any vertex of $\Rcal$, that face
		must be a triangle ($K_3$).
		Hence no such $C$ exists and $D(\Rcal)$ is chordal.
		\qed
	\end{proof}
	
	A vertex-guard $g \in V(\Rcal)$ covers exactly its incident rectangles $C(g)$, which form a clique of size at most three in $D(\Rcal)$.
	Conversely, every clique of $D(\Rcal)$ has size at most three and is contained in some $C(v)$: a $K_3$ corresponds to a T-junction, a $K_2$ not contained in a $K_3$ corresponds to a boundary vertex $v$ with $|C(v)| = 2$, and a maximal $K_1$ corresponds to an outer corner of $\Omega$.
	Thus, \textsc{RVG} reduces to \textsc{Minimum Clique Cover} on $D(\Rcal)$.
	Since $D(\Rcal)$ is chordal by~\Cref{lemma:boundary_exposed_rectangulation_dual_graph_chordal_outerplanar}, it is perfect and admits a \emph{perfect elimination ordering} (PEO), from which a minimum clique cover can be obtained greedily.

	\begin{theorem}
		\textsc{RVG} can be solved in $\bigO(n)$ time for boundary-exposed rectangulations with $n$ rectangles.
	\end{theorem}
	
	\begin{proof}
		Because $|V(\Rcal)| = 2n+2$ and each $v \in V(\Rcal)$ contributes at most $\binom{3}{2} = 3$ edges $\{r_i, r_j\} \subseteq C(v)$ to $E_D$,
		we can construct $D(\Rcal)$ from the standard representation in $\bigO(n)$ time.
		A PEO $\sigma$ of $D(\Rcal)$ can be computed by Maximum Cardinality Search in $\bigO(|V_D| + |E_D|) = \bigO(n)$ time~\cite{rose1976algorithmic}, and the greedy clique-cover algorithm of Gavril~\cite{gavril1972algorithms} scans $\sigma$ and $D(\Rcal)$ once, in $\bigO(n)$ time as well (see also~\cite{rose1976algorithmic}).
		Specifically, if $\sigma = v'_1, \dots, v'_n$ is a PEO, each later closed neighborhood $N^+[v'_i] = \{v'_i\} \cup (N(v'_i) \cap \{v'_{i+1}, \dots, v'_n\})$ is a clique.
		We process the vertices in the order $\sigma$ and, whenever a vertex $v'_i$ is still uncovered, place a corresponding guard to cover $N^+[v'_i]$.
		This produces a maximum independent set, which in a perfect graph has the same size as a minimum clique cover.
		\qed
	\end{proof}

	\subsection{Generalization to Few Interior Rectangles}
	\label{sec:boundary_exposed_rectangulations_generalization}
	
	In this section we prove~\Cref{thm:fpt_algorithm}, which extends our approach to rectangulations that are \emph{almost} boundary-exposed, i.e., where few \emph{interior rectangles} are completely surrounded by others, thereby solving a much wider family of instances efficiently.
	
	\begin{theorem}
		\label{thm:fpt_algorithm}
		Let $\Rcal$ be a rectangulation of $\Omega$ with $n$ rectangles, exactly $k$ of which do not touch $\bd \Omega$.
		Then \textsc{RVG} on $\Rcal$ is fixed-parameter tractable parameterized by $k$, with running time
		$2^{\bigO(k)} \cdot n$.
	\end{theorem}
	
	This parameterization essentially interpolates between the linear case $k = 0$ (\Cref{sec:boundary_exposed_rectangulations_linear_time}) and the hard case $k \in \Theta(n)$ (\Cref{sec:rectangulations_np_completeness}), thus providing a middle ground with exponential dependence only on the distance to boundary-exposure.
	
	Again, fix a planar embedding of $D(\Rcal)$ in which a rectangle touches $\bd \Omega$ if and only if its dual vertex is incident to the outer face, and let $I \subseteq V_D$ be the $k$ interior rectangles.
	The key idea is then to exploit $I$ being an \emph{outerplanar modulator}: deleting $I$ leaves an outerplanar graph.
	
	\begin{lemma}
		\label{lemma:boundary_exposed_rectangulation_treewidth_dual_graph_bounded}
		$\tw(D(\Rcal)) \le k+2$.
	\end{lemma}
	
	\begin{proof}
		It is well known that $\tw(G) \le \tw(G \setminus S) + |S|$ for any graph $G$ and $S \subseteq V(G)$.
		Deleting $I$ from the fixed embedding leaves an outerplanar graph, so $\tw(D(\Rcal) \setminus I) \le 2$.
		Take a tree decomposition $(T, \Xcal)$ of $D(\Rcal) \setminus I$ of width at most~$2$ and add all vertices of $I$ to every bag ($X'_t = X_t \cup I$ for every $t \in V(T)$).
		The resulting structure satisfies all three axioms, so it yields a tree decomposition of $D(\Rcal)$ whose bags have size at most $k+3$, thus $\tw(D(\Rcal)) \le k+2$.
		\qed
	\end{proof}

	The weak dual graph $D(\Rcal)$ is not expressive enough by itself to encode guard placement, so we lift this decomposition to the bipartite
	graph $H(\Rcal) = (V_H, E_H)$ used in~\Cref{sec:approximability_and_ptas}:
	since $H(\Rcal)$ is planar, \textsc{RVG} reduces to \textsc{PRBDS} on this graph.
	We now prove that the lifting preserves a bound on the treewidth of $H(\Rcal)$.
	
	\begin{lemma}
		\label{lemma:boundary_exposed_rectangulation_treewidth_incidence_graph_bounded}
		If $k \ge 1$, $\tw(H(\Rcal)) \le k + 2$.
	\end{lemma}
	
	\begin{proof}
		Let $(T, \Xcal')$ be a tree decomposition of $D(\Rcal)$ of width at most $k+2$ as guaranteed by~\Cref{lemma:boundary_exposed_rectangulation_treewidth_dual_graph_bounded}.
		For every $v \in V(\Rcal)$, the incident rectangles $C(v)$ form a clique of size at most three in $D(\Rcal)$, and it is a well-known property of tree decompositions that every clique is contained in some bag $X'_{t_v}$.
		Append to any such $t_v$ a new leaf node $t'_v$ with bag $X'_{t'_v} = C(v) \cup \{v\}$.
		The resulting structure is a tree decomposition of $H(\Rcal)$: every rectangle already appears in $(T, \Xcal')$, every vertex $v$ appears in its leaf bag $X'_{t'_v}$, and every edge $\{r, v\}$ of $H(\Rcal)$ satisfies $r \in C(v)$ and is therefore covered by the leaf bag $X'_{t'_v} = C(v) \cup \{v\}$.
		Moreover, the original bags containing a rectangle $r$ form a connected subtree of $T$, and each new leaf $t'_v$ added for a vertex $v$ incident to $r$ is attached to a node $t_v$ inside that subtree, preserving connectedness; similarly, every vertex $v \in V(\Rcal)$ appears in a single leaf bag, which is trivially connected.
		The original bags have size at most $k + 3$,
		while each new leaf bag has size $|C(v)| + 1 \le 4$,
		hence $\tw(H(\Rcal)) \le \max\{k+3, 4\} - 1 \le k + 2$.
		\qed
	\end{proof}

	Since $\tw(H(\Rcal)) \le k+2$, we can construct a nice tree decomposition of $H(\Rcal)$ of width $\bigO(k)$ and $\bigO(k \cdot n)$ bags in $2^{\bigO(k)} \cdot n$ time~\cite{cygan2015parameterized,kloks1994treewidth}.
	Applying the dynamic programming algorithm of~\Cref{sec:ptas_treewidth_dynamic_programming} to the resulting \textsc{PRBDS} instance takes $2^{\bigO(k)}$ time per bag, giving a total of $\bigO(2^{\bigO(k)} \cdot k \cdot n) = 2^{\bigO(k)} \cdot n$.
	Hence, \textsc{RVG} is FPT parameterized by the number of interior rectangles $k$.

	\paragraph{Open questions.}
	Several research lines remain open.
	An interesting direction is to investigate the existence of a \emph{below-guarantee parameterization} by the difference $\delta = \lceil n/2 \rceil - \mathrm{OPT}$, which measures the gap between the known upper bound $\lceil n/2 \rceil$
	and the optimum.
	While our hole-filling gadget $\Phi$ corresponds to a guillotine partition, neither hardness nor polynomial-time tractability for general guillotine rectangulations follows immediately from our results, though we conjecture that \textsc{RVG} can be solved in polynomial time for this class.

	\begin{credits}
		\subsubsection{\ackname}
		We thank the anonymous reviewers for their valuable feedback.
		This work was partially supported by CMUP, member of LASI, which is financed by national funds through FCT -- Fundação para a Ciência e a Tecnologia, I.P., under the project with DOI 10.54499/UID/00144/2025.
		Also supported by the URBAN-NET project, with reference NORTE2030-FEDER-02697300, co-financed by the European Union through the NORTE 2030 Regional Program, under Portugal~2030.
		
		\subsubsection{\discintname}
		The authors have no competing interests to declare that are
		relevant to the content of this article.
	\end{credits}
	\bibliographystyle{splncs04}
	\bibliography{refs.bib}

\end{document}